# An Interline Multiport DC Circuit Breaker with Power Flow Controller

**Maliha FERDOUS**
**The University of British Columbia**
**Canada**
**malihaferdous@ece.ubc.ca**

**Zhi Jin ZHANG***
**The University of British Columbia**
**Canada**
**justin.zhang@ubc.ca**

**Qichen YANG**
**University of Central Florida**
**USA**
**qichen.yang@ucf.edu**

**Maryam SAEEDIFARD**
**Georgia Institute of Technology**
**USA**
**maryam@ece.gatech.edu**

## SUMMARY

Currently, the high-voltage multi-terminal DC (MTDC) grids face two major challenges. First, fast rise rate and lack of zero-crossing point in DC-side fault currents render AC circuit breakers inoperable. Second, DC line currents are passively determined by terminal DC voltages and line resistances, which inherently limits the capability of changing individual line currents for essential grid services such as power flow management and overloading prevention. To address these two limitations, DC circuit breakers (DCCBs) and DC power flow controllers (DCPFCs) have been developed. The conventional DCCBs are two-port devices installed in series with DC lines at their terminals in an MTDC converter substation, so that any fault current flowing through the line can be rapidly cleared. Additionally, a DCPFC can be installed at line terminals to adjust terminal voltages and power exchanges among adjacent lines. As a result, for a substation that connects to multiple lines, it needs to host multiple DCCBs and a DCPFC, which significantly increases the cost and footprint of both the substation and the MTDC grid. To this end, this work proposes an integrated equipment combining multiple DCCBs and a DCPFC, called the interline multiport DCCB with DCPFC, which can reduce the number of components and enhance the economic feasibility of MTDC networks. This paper presents the topology, operation principles, and component selection procedures of the proposed equipment, and compares it with the state-of-the-art technologies. Electromagnetic transient (EMT) simulation results validate the fault interruption and power flow control performance of the interline multiport DCCB with PFC.

## KEYWORDS

Multi-terminal DC (MTDC), high-voltage DC (HVDC), DC circuit breaker (DCCB), DC power flow controller (DCPFC), multiport, interline.

# 1 Introduction

High-voltage multi-terminal DC (MTDC) transmission systems based on voltage-sourced converters are regarded as a key enabler for the large-scale integration of renewable energy sources and the modernization of electric power grid. However, MTDC grids face two major technical barriers. First, unlike AC fault current, DC fault current rises sharply and does not possess a zero-crossing point. Second, power flows in DC lines are determined passively by line resistances and the node voltages, which may lead to uncontrolled power flow scenarios that overload or under-utilize the lines. In this context, DC circuit breakers (DCCBs) and DC power flow controllers (DCPFCs) have been developed to address the aforementioned barriers, respectively [1]. The state-of-the-art high-voltage DCCB is a two-port device consisting of a fast mechanical switch (FMS), solid-state switches (typically IGBTs), and metal-oxide varistors (MOVs). The DCCB is installed at the terminal of each line to ensure quick and selective fault interruption performance. On the other hand, the DCPFC can be divided into resistance- or voltage-based devices, which vary the resistance or the voltage of the lines, respectively. The voltage-based DCPFCs are generally preferred as they lead to lower losses compared to their resistance-based counterparts, and the variable voltages are provided by power electronic converters. Therefore, the existing DCPFCs can be regarded as multiport converters that connect to multiple transmissions line to modulate their current flows and power sharing [2].

From the discussions above, it can be observed that for each MTDC converter substation that connects to $n$ lines, it has to host $n$ DCCBs and a DCPFC (as shown in Fig. 1(a)), which exacerbates the substation cost and footprint. In light of this, the integration of multiple DCCBs and a DCPFC into a single piece of equipment is envisioned to reduce the component count and enhance the MTDC grid's practical feasibility. This concept of an integrated multiport DCCB with DCPFC has received attention in recent years. To transform from two-port into multiport DCCBs, most of the state-of-the-art designs create topologies that share a common energy absorption module (highlighted with a red box in Fig. 1(b)) across all lines [3]. This is combined with commutation circuits that can route the fault current from any line to the shared module. The energy absorption module, composed of high-voltage IGBTs and MOVs, is responsible for dissipating the fault energy; as a result, it is costly and bulky. Therefore, by reducing the number of these modules, which would otherwise present in every two-port DCCBs, into a single unit, the multiport DCCB's cost and size can be significantly decreased [3]. Subsequently, to integrate multiport DCCB with DCPFC, half-bridge modules, full-bridge modules, and customized circuitry have been developed to simultaneously serve the role of fault current commutation circuit and current flow controller [4-6]. During PFC operations, the lines are capacitively coupled, and the capacitors enable voltage adjustments and power exchanges.

However, the state-of-the-art designs combining multiport DCCB with DCPFC suffer one or more of the following drawbacks. First, when faults happen on multiple lines, one line fault must be completely cleared until the next fault can be interrupted. This is because the single energy absorption module prevents simultaneous tripping. Second, when the energy absorption module malfunctions, potentially due to issues such as compromised tripping signals, and/or broken IGBTs, IGBT gate drivers, MOVs, etc., the multiport DCCB cannot interrupt any fault, owing to the single point of failure and the lack of redundancy. Moreover, the DCPFC integration involves a considerable number of bridge-based or specialized switching cells, which are complex to control and operate in conjunction with the DCCB components, resulting in an increased number of active semiconductor switches and intricate tripping sequences.

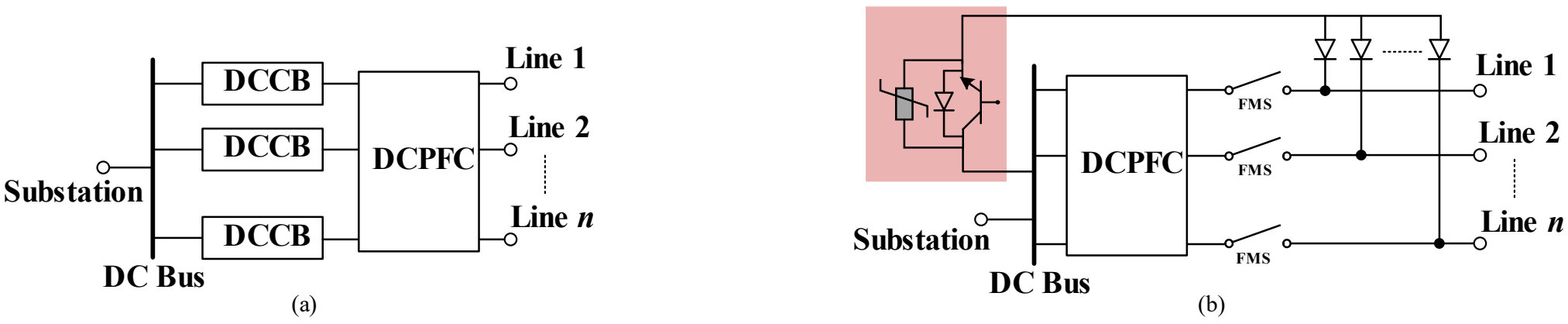

***Figure 1 – (a) The conventional two-port DCCB and DCPFC installation scheme (note that each two-port DCCB contains an energy absorption module), (b) A representative state-of-the-art multiport DCCB integrated with DCPFC topology (the shared energy absorption module is highlighted with a red box).***

To bridge the above-mentioned gaps, this work introduces a multiport DCCB with DCPFC capability in an interline configuration (denoted as **IM-PFCCB**), by integrating the interline DCCB [4] and DCPFC [1,7] topologies. The IM-PFCCB, although still a concept at the research stage, provides a promising trade-off between the conventional scheme and the existing multiport DCCB designs, as it (i) allows the simultaneous tripping of multiple lines, (ii) offers redundancy, and (iii) incorporates a simple DCPFC. The fault interruption and power flow control strategies, as well as component selection procedures are presented. The efficacy of the proposed IM-PFCCB is verified via electromagnetic transient (EMT) simulation results.

## 2 Topology of the Proposed IM-PFCCB

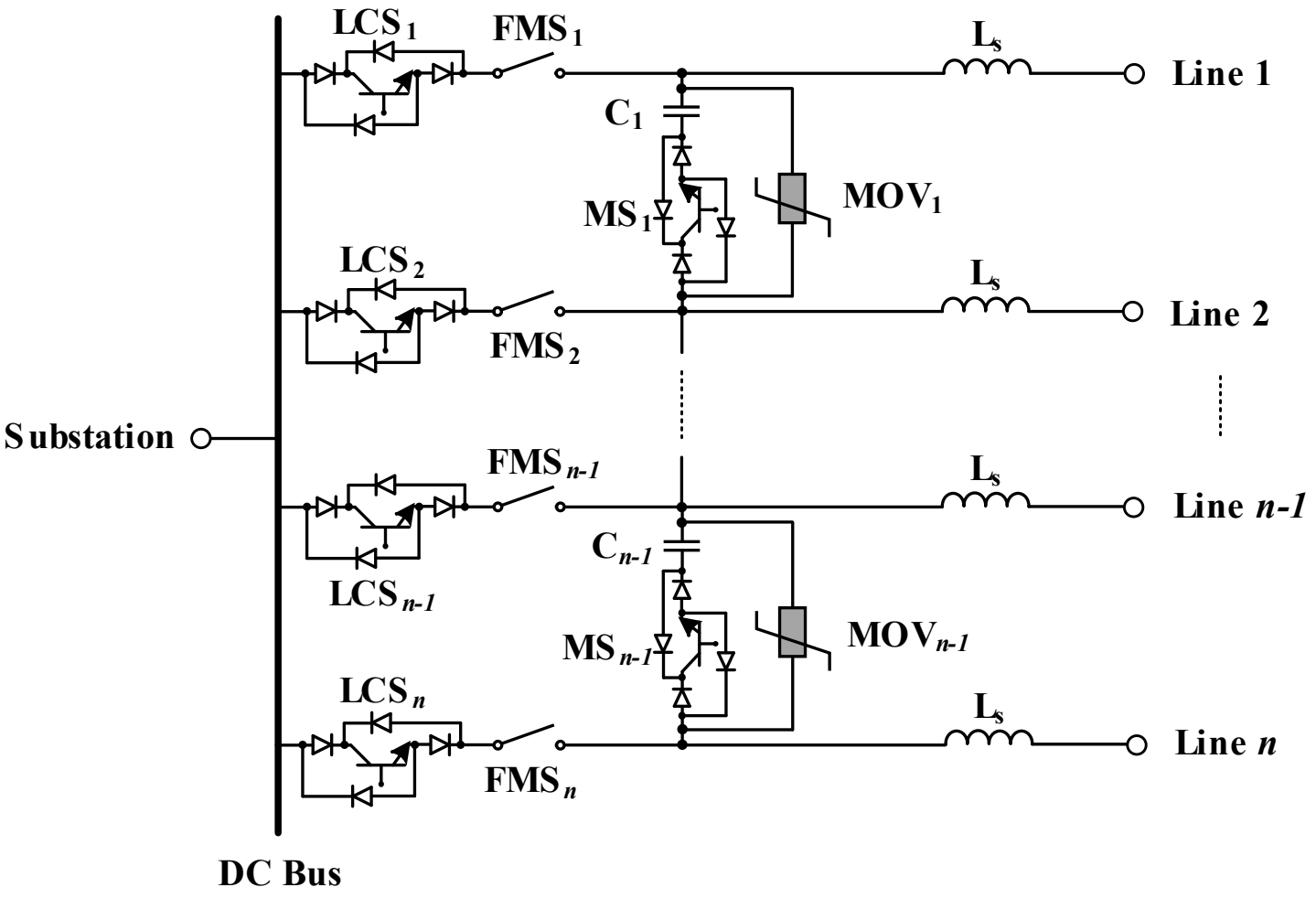

***Figure 2 – The proposed IM-PFCCB.***

The schematic of the IM-PFCCB is shown in Fig. 2. The proposed device has $n + 1$ ports, which connect to $n$ lines and a converter substation. Each line connects to the DC bus via a current-limiting inductor ($L_S$), an FMS, and a load commutation switch (LCS). To enable bidirectional fault interruption and power flow control, rather than connecting two IGBTs back-to-back in series, the LCS is realized via one IGBT and four diodes, since diodes are generally cheaper and easier to operate compared to IGBTs [1]. Connecting between each line are two parallel branches (i.e., the interline components) – one branch contains the MOV and another branch contains a capacitor (C) in series with a main switch (MS). The MS is realized using the same IGBT/diode connections as the LCS for bidirectional switching. The MOV, C, and MS are responsible for fault energy absorption, power flow control, and fault current commutation, respectively. Overall, the topology presented in Fig. 2 depicts that for an IM-PFCCB with $n+1$ ports, it has $n-1$ energy absorption modules. This number can be further reduced at the expense of operational flexibility and redundancy, which is discussed in Section 6.1.

# 3 Operation Principles of the Proposed IM-PFCCB

## 3.1 Line Fault Interruption

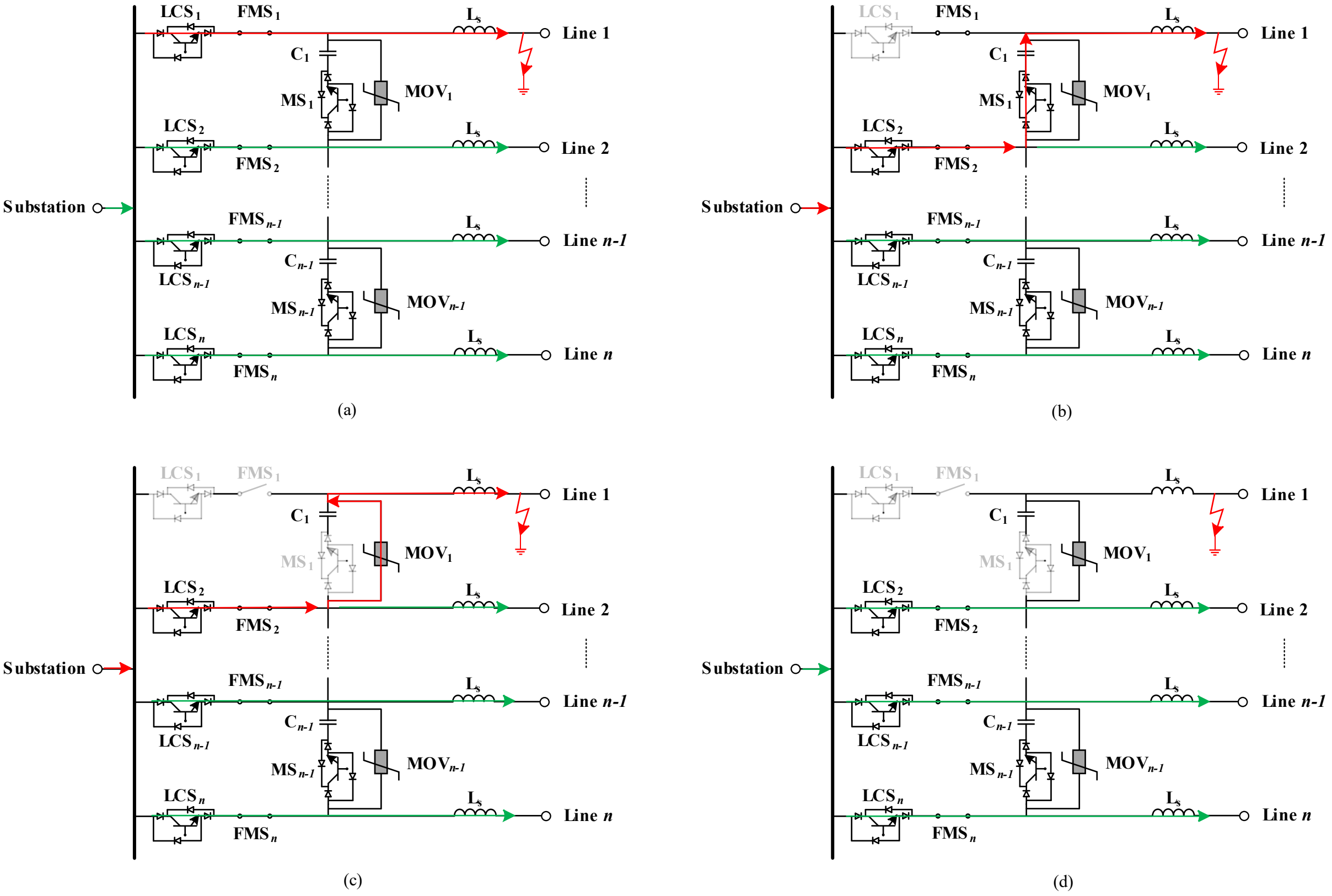

*Figure 3 – Fault current path: (a) $t_0 - t_2$, (b) $t_2 - t_3$, (c) $t_3 - t_4$, (d) $t \geq t_4$.*

The fault current path in IM-PFCCB is illustrated in Fig. 3. Bidirectional fault currents can be interrupted by this device. During normal operations (before time $t_0$), all FMS are closed. The IM-PFCCB can operate in either (i) PFC mode, where all MS are closed and LCS are switching to regulate line currents, or (ii) non-PFC mode, where all MS are open and LCS are closed, and the line currents are purely determined by line resistances and converter terminal voltages.

At $t_0$, assuming a fault occurs on Line 1 (Fig. 3(a)). Subsequently, the fault is detected at time $t_1$. At this instant, the LCS on Line 1 (i.e., $LCS_1$) is ordered to open and $MS_1$ is ordered to close. As a result, the fault current is commutated from the $LCS_1$ – $FMS_1$ path to the interline $MS_1$ – $C_1$ path, which is completed at time $t_2$ (Fig. 3(b)). At the same time, $FMS_1$ starts to open. Once $FMS_1$ has gained sufficient voltage withstand capability at $t = t_3$, $MS_1$ is turned off, which routes the fault current into $MOV_1$ (Fig. 3(c)). It is also possible to charge $C_1$ to the clamping voltage of $MOV_1$ so that the fault current is naturally commutated to $MOV_1$. This allows the use of thyristor, rather than IGBT, in the MS. However, it requires $C_1$ to have a high voltage rating, which is not desirable. Once the fault current is commutated to $MOV_1$, it dissipates the fault energy. At time $t_4$, the fault is extinguished and Line 1 can be isolated (Fig. 3(d)). During the Line 1 fault removal process, current flows on other lines are not impacted.

It can be observed that due to the interline MOV placements, multiple line faults can be simultaneously interrupted by IM-PFCCB. For example, if faults occur on both Line 1 and Line 2, then $MOV_1$ can be used to break the fault on Line 1, and $MOV_2$ (the interline MOV between Line 2 and Line 3) can be used to interrupt Line 2 fault, as depicted by Fig. 4(a), following the exact operations outlined above with $LCS_2$, $FMS_2$, and $MS_2$.

Lastly, protection redundancy is illustrated in Fig. 4(b). Assume a fault occurs on Line 1, but $MOV_1$ has degraded and cannot perform fault interruption, and/or $MS_1$ cannot be turned off due to gate driver or tripping signal propagation issues. Then $LCS_1$, $FMS_1$, $LCS_2$, and $FMS_2$ are all ordered to open, commutating the fault current into the interline branches associated with $MS_1$ and $MS_2$. Subsequently, $MS_2$ is turned off, routing the fault current into $MOV_2$, which performs fault interruption for Line 1.

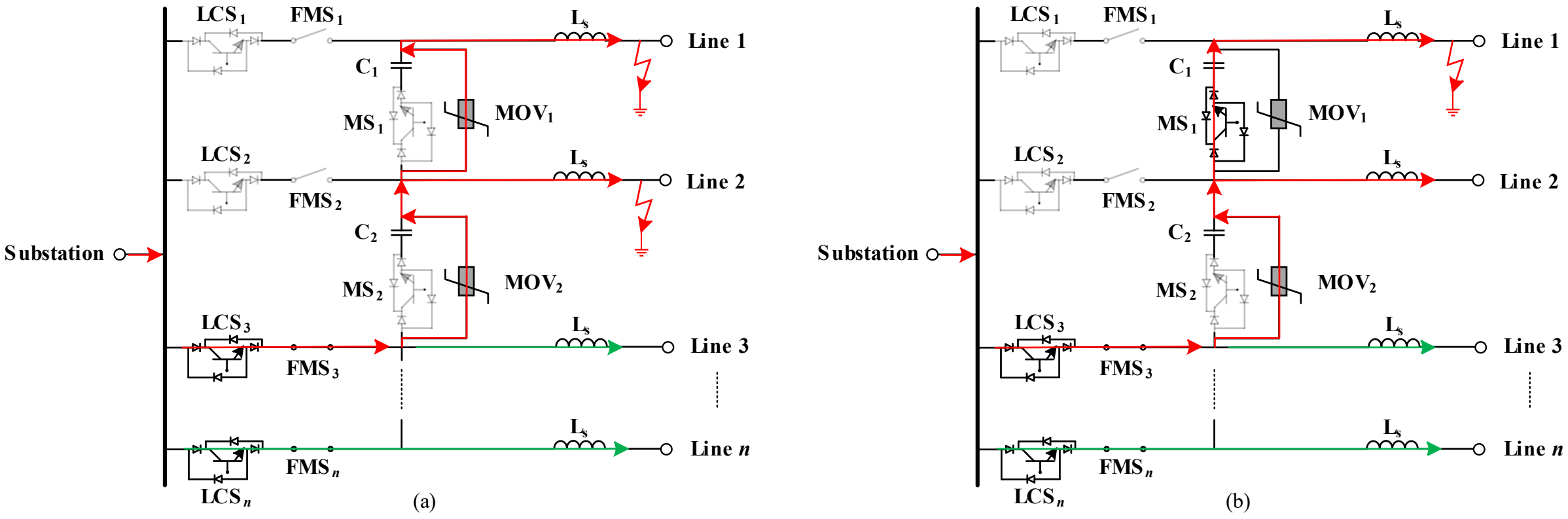

***Figure 4 – Fault current path: (a) simultaneous fault interruption on Line 1 and Line 2, (b) Line 1 fault interruption using $MOV_2$ in case $MOV_1$ is degraded and/or $MS_1$ does not turn off.***

## 3.2 Power Flow Control

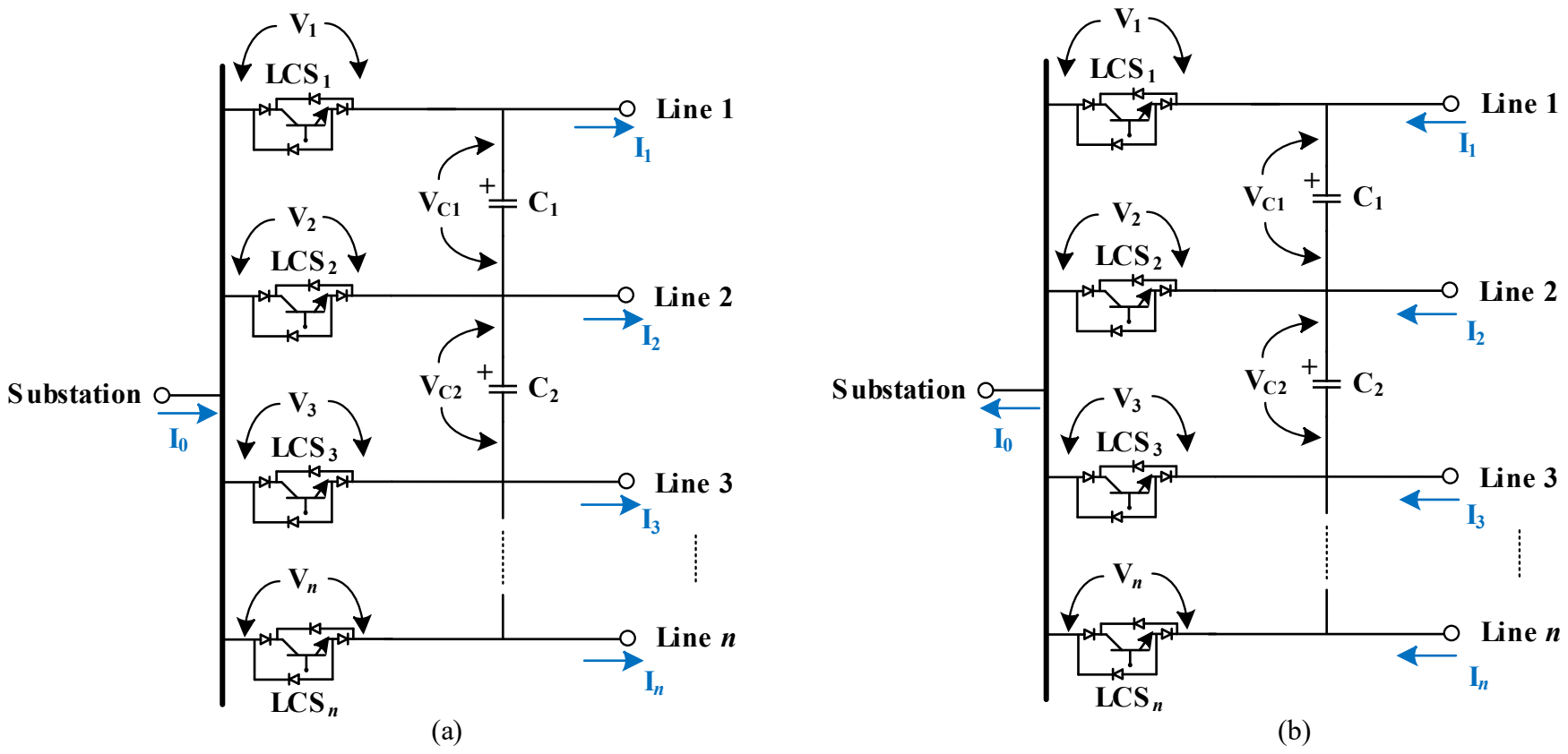

***Figure 5 – PFC scenarios: (a) Current flowing out of substation, (b) Current flowing into substation.***

During the PFC mode, all FMS and MS remain closed and the IM-PFCCB can be simplified as Fig. 5, where each line has an active switch (LCS) and coupled by the interline capacitors. The PFC mode only works when all line currents are in the same direction (either into or out of the DC bus), so there are only two possible working conditions as shown in Figs. 5(a) and 5(b), respectively. The proposed scheme simplifies PFC operation and is highly scalable, as it can be readily extended to multiple lines. By contrast, accommodating mixed current directions requires additional circuitry and control coordination, resulting in a more complex design and reduced practicality [7-8].

The principles of PFC operation are explained based on the scenario in Fig. 5(a); they are equally applicable to the scenario in Fig. 5(b). During PFC operations, the $\mathrm{LCS}_i$ $(i = 1, \dots, n)$ are turned on and off repetitively with a switching frequency of $f_{sw}$ (typically less than 10 kHz for IGBTs). During a switching period $T_{sw} = 1/f_{sw}$, only one $\mathrm{LCS}_i$ is turned on at any time with a duration determined by its duty cycle $D_i$. This is expressed by

$$\sum_{i=1}^{n} t_{LCS_i} = T, \tag{1}$$

$$\sum_{i=1}^{n} D_i = 1, \tag{2}$$

where $t_{LCS_i}$ is the ON-time of $\mathrm{LCS}_i$ and $D_i = t_{LCS_i}/T$. Assuming the line currents remain constant during $T_{sw}$, then currents through capacitors $\mathrm{C}_j$ $(j = 1, \dots, n-1)$, can be derived as

$$I_{C_j} = \sum_{k=j+1}^{n} I_k, \qquad \text{for } i \le j \tag{3}$$

$$I_{C_j} = \sum_{k=1}^{j} (-I_k), \qquad \text{for } i > j \tag{4}$$

where subscript $i$ denotes the $\mathrm{LCS}_i$ that is turned on. Since the capacitor current must be balanced, then

$$\sum_{i=1}^{n} D_i I_{C_j} = 0. \tag{5}$$

Substituting (3) and (4) into (5) to obtain

$$\sum_{i=1}^{j} D_i \sum_{k=j+1}^{n} I_k + \sum_{i=j+1}^{n} D_i \sum_{k=1}^{j} (-I_k) = 0. \tag{6}$$

Subsequently, substituting (2) into (6) to derive

$$\left(1 - \sum_{i=j+1}^{n} D_i\right) \sum_{k=j+1}^{n} I_k + \sum_{i=j+1}^{n} D_i \sum_{k=1}^{j} (-I_k) = 0,$$

$$\sum_{k=j+1}^{n} I_k - \sum_{i=j+1}^{n} D_i \sum_{k=1}^{n} I_k = 0.$$

As a result,

$$\sum_{i=j+1}^{n} D_i = \frac{\sum_{k=j+1}^{n} I_k}{\sum_{k=1}^{n} I_k} = \frac{\sum_{k=j+1}^{n} I_k}{I_0}. \tag{7}$$

For individual $D_i$, it is possible to obtain

$$D_i = \sum_{k=i}^{n} D_k - \sum_{k=i+1}^{n} D_k = \frac{\sum_{k=i}^{n} I_k - \sum_{k=i+1}^{n} I_k}{I_0}.$$

Hence,

$$D_i = \frac{I_i}{I_0}. \tag{8}$$

Based on (8), the current flow on Line $i$ is proportional to the duty cycle of its $\mathrm{LCS}_i$. Therefore, by applying the desired $D_i$ to $\mathrm{LCS}_i$ $(i = 1, \dots, n-1)$, the individual power flows on Lines 1 to $n-1$ can be adjusted, whereas the flow on Line $n$ is passively determined as a result of Kirchhoff's Current Law (i.e., $\sum_{i=1}^{n} I_i = I_0$). A closed-loop control strategy can also be used, where reference current values for Lines 1 to $n-1$ are subtracted by their respective measured currents and the resulting errors are passed into proportional-integral controllers to issue duty cycles $D_i$ $(i = 1, \dots, n-1)$ [7]. Lastly, the capacitor voltages $V_{C_i}$ $(i = 1, \dots, n-1)$ depends on the current flows and converter voltages [7-8]. As these capacitor voltages set the maximum voltage that can be imposed by the PFC, their ratings directly limit the PFC operating range. In practice, the capacitor voltages are set to be a few percent of the rated DC bus voltage [7-8].

## 4 Component Selection Procedures

The selection guidelines for LCS, MOV, MS, and interline capacitor are provided. To meet the required voltage and current ratings and to scale to higher operating levels, multiple devices may be connected in series and parallel as needed [8-9].

*1) LCS:* During the fault interruption mode, the LCS should tolerate the maximum anticipated fault current due to the conduction of the fault current in adjacent faulty lines. The maximum fault current can be calculated using well-established formula [9]. This sets the current rating of the semiconductor devices. On the other hand, LCS does not need to block MOV voltage (which is imposed on the FMS), and its blocking voltage is determined by the sum of capacitor voltages between the LCS in the ON state and the LCS in the OFF state during PFC mode. Thus, the maximum blocking voltage is the sum of all capacitor voltages:

$$V_{LCS} = \sum_{k=1}^{n-1} V_{C_k}. \tag{9}$$

Since the IGBT and diode are in parallel during the LCS' OFF state, their voltage ratings have to be $k_{sf} \times V_{LCS}$, where $k_{sf} > 1$ is a safety margin factor.

*2) MOV:* The MOV is responsible for fault energy dissipation and can be selected based on well-established procedures [9].

*3) MS:* The current rating of the IGBT and diode in MS is set by the maximum fault current, identical to LCS. Additionally, the MS needs to block the MOV clamping voltage ($V_{MOV}$). Therefore, the voltage rating of the MS' IGBT and diode can be selected as $k_{sf} \times V_{MOV}$.

*4) Capacitor:* As shown in Section 3.2, the capacitor voltage rating is application-specific, but it can be much lower than $V_{MOV}$. The capacitance can be selected based on the maximum permitted voltage ripple and is derived as follows. For capacitor $C_j$ ($j = 1, \dots, n-1$), its voltage change ($\Delta V_{Cj}$) during the ON-time of LCS$_i$ is

$$\Delta V_{C_j} = \frac{I_{C_j} D_i T_{sw}}{C_j}. \tag{10}$$

Therefore, for all $i \leq j$, $C_j$ is charging and $\Delta V_{Cj}$ can be expressed as

$$\Delta V_{C_j,charge} = \sum_{k=j+1}^{n} I_k \sum_{i=1}^{j} D_i \frac{T_{sw}}{C_j} = \sum_{k=j+1}^{n} I_k \frac{\sum_{k=1}^{j} I_k}{I_0} \frac{T_{sw}}{C_j}. \tag{11}$$

On the other hand, for all $i > j$, $C_j$ is discharging and $\Delta V_{Cj}$ can be expressed as

$$\Delta V_{C_j,discharge} = \sum_{k=1}^{j} I_k \sum_{i=j+1}^{n} D_i \frac{T_{sw}}{C_j} = \sum_{k=1}^{j} I_k \frac{\sum_{k=j+1}^{n} I_k}{I_0} \frac{T_{sw}}{C_j}. \tag{12}$$

It can be seen that (11) and (12) are equivalent and the capacitor voltage is balanced during steady-state operations. Based on (11) and (12), the capacitance can be selected as

$$C_j \geq \frac{\left(\sum_{k=1}^{j} I_k\right)\left(I_0 - \sum_{k=1}^{j} I_k\right)}{f_{sw}\, I_0\, \Delta V_{C_j,max}}, \tag{13}$$

where $\Delta V_{C_j,max}$ is the maximum permitted capacitor voltage ripple.

# 5 Simulation Results

The efficacy of IM-PFCCB is verified on the four-terminal 320 kV MTDC grid benchmark EMT model [10]. Detailed system information is omitted here due to space limitations and can be found in [10]. IM-PFCCB is placed at a converter substation connected to three lines, which are labelled as Line 1, 2, 3, respectively. The substation is rated at 900 MVA. All line currents are made to flow out of the converter substation to satisfy the same-direction current requirement for DCPFC operation. Using procedures outlined in Section 4, the MOV clamping voltage $V_{MOV}$ is selected to be 600 kV, the switching frequency $f_{sw}$ is set at 1 kHz, and the capacitor is chosen as 1500 $u$F to keep a maximum voltage ripple at 5% with a rated capacitor voltage of 5 kV (i.e., maximum ripple is 250 V).

## 5.1 Fault Interruption Performance

First, the capability of simultaneously interrupting multiple faults, as discussed in Fig. 4(a), is demonstrated. In this scenario, a bolted Line 1 fault occurs at $t = 2$ sec. and a bolted Line 2 fault occurs at $t = 2.001$ sec. Note that the faults can happen at exactly the same time, but the resulting waveforms will have significant overlap, so Line 2 fault is delayed by 1 ms for illustrative clarity. It is assumed that the fault detection time is 1 ms, and FMS takes 2 ms to open. Figures 6(a) and 6(b) show all line voltages and currents, respectively, during Line 1 and Line 2 faults. They demonstrate that both faults are successfully cleared within 5 ms and 9 ms since their inception, respectively, while the healthy Line 3 voltage and current are not affected.

Figure 6(c) depict the MOV voltages. It can be observed that when the fault currents are routed into $MOV_1$ and $MOV_2$ at $t = 2.003$ sec. and $t = 2.004$ sec., respectively, the MOV voltages increase from zero to $V_{MOV}$. Once Line 1 fault current is completely extinguished at $t = 2.005$ sec., $MOV_1$ voltage decreases to around $-270$ kV. This results from the source voltage (320 kV) being subtracted by the $MOV_2$ voltage (590 kV) during Line 2 fault interruption, while Line 1 current is already zero. Overall, $MOV_1$ and $MOV_2$ dissipate 3 MJ and 6.5 MJ of energy during the fault interruption, respectively.

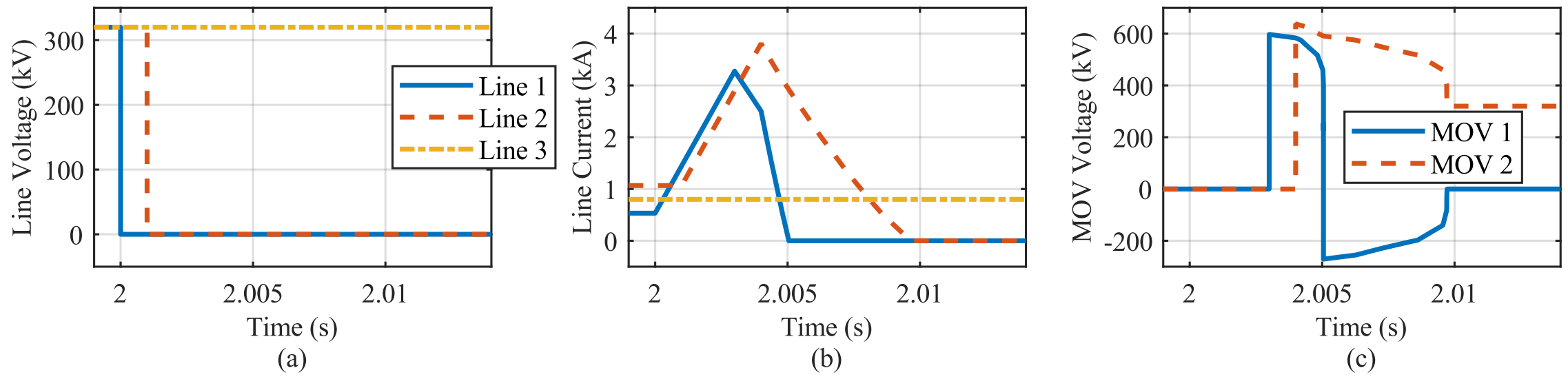

***Figure 6 – Multiple line tripping: (a) Line voltages, (b) Line currents, (c) MOV voltages.***

Next, protection redundancy is illustrated. In this hypothetical scenario, a bolted Line 1 fault occurs at $t = 2$ sec., but $MOV_1$ cannot break the fault due to degradation (or $MS_1$ is unable to turn off due to gate driver/signal issues). As a result, the backup fault interruption sequence shown in Fig. 4(b) is activated at 1 ms after fault detection. The line voltages, currents, and MOV voltages are depicted in Figs. 7(a), 7(b), and 7(c), respectively. It is observed that even though $MOV_1$ does not participate in Line 1 fault clearing, the fault is routed into $MOV_2$ at $t = 2.004$ sec. and interrupted at $t = 2.01$ sec., thanks to the redundancy provided by $MOV_2$. The overall fault clearing time is 10 ms, and the fault energy absorbed by $MOV_2$ is approximately 7 MJ. Although Line 2 current is also interrupted, it can be resumed as soon as the faulty Line 1 is isolated (e.g., by a residual current switch).

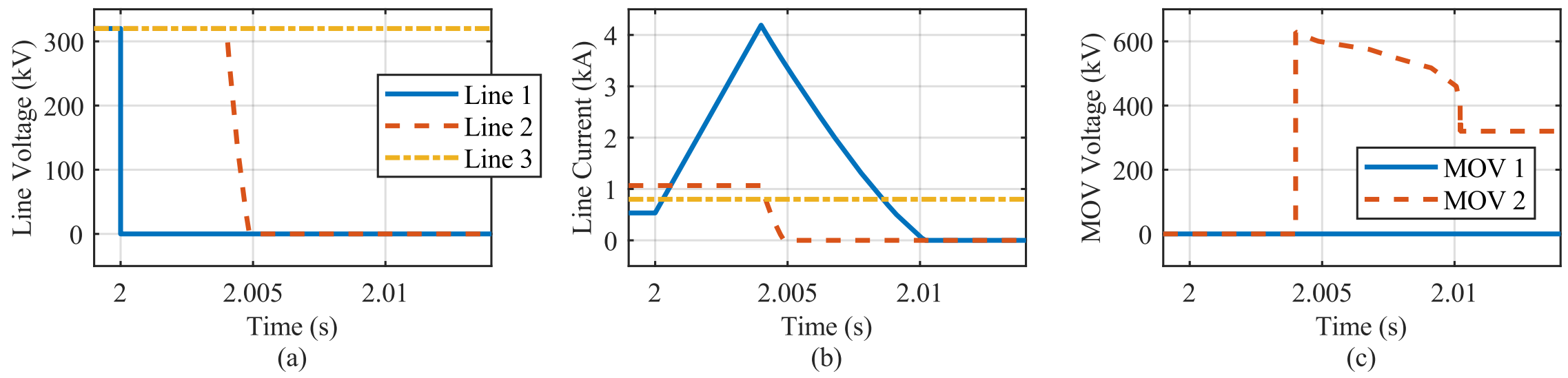

***Figure 7 – Using $MOV_2$ to trip Line 1 fault: (a) Line voltages, (b) Line currents, (c) MOV voltages.***

## 5.2 PFC Performance

A closed-loop controller is implemented to control Line 1 and Line 2 currents. Initially, the converter outputs rated power and all three lines have a current flow of 1 kA. At $t = 5$ sec., the DCPFC is turned on and the reference currents for Line 1 and Line 2 are set to 1.3 and 0.75 kA, respectively. Subsequently, at $t = 25$ sec., Line 1 current reference is changed to 0.9 kA. Figure 8(a) demonstrates that Line 1 current closely tracks its set-point. Figure 8(b) shows all three line currents during PFC operation. It is observed that Line 2 current stays at its reference value of 0.75 kA, and Line 3 current changes accordingly based on Kirchhoff's Current Law.

Figure 8(c) depicts the capacitor voltage waveforms, which are kept within their rated voltage of 5 kV. Fig. 8(d) shows that the capacitor voltage ripple is 230 V, which is less than its maximum permitted value of 250 V. This verifies the functionality of DCPFC and the proposed component selection procedures.

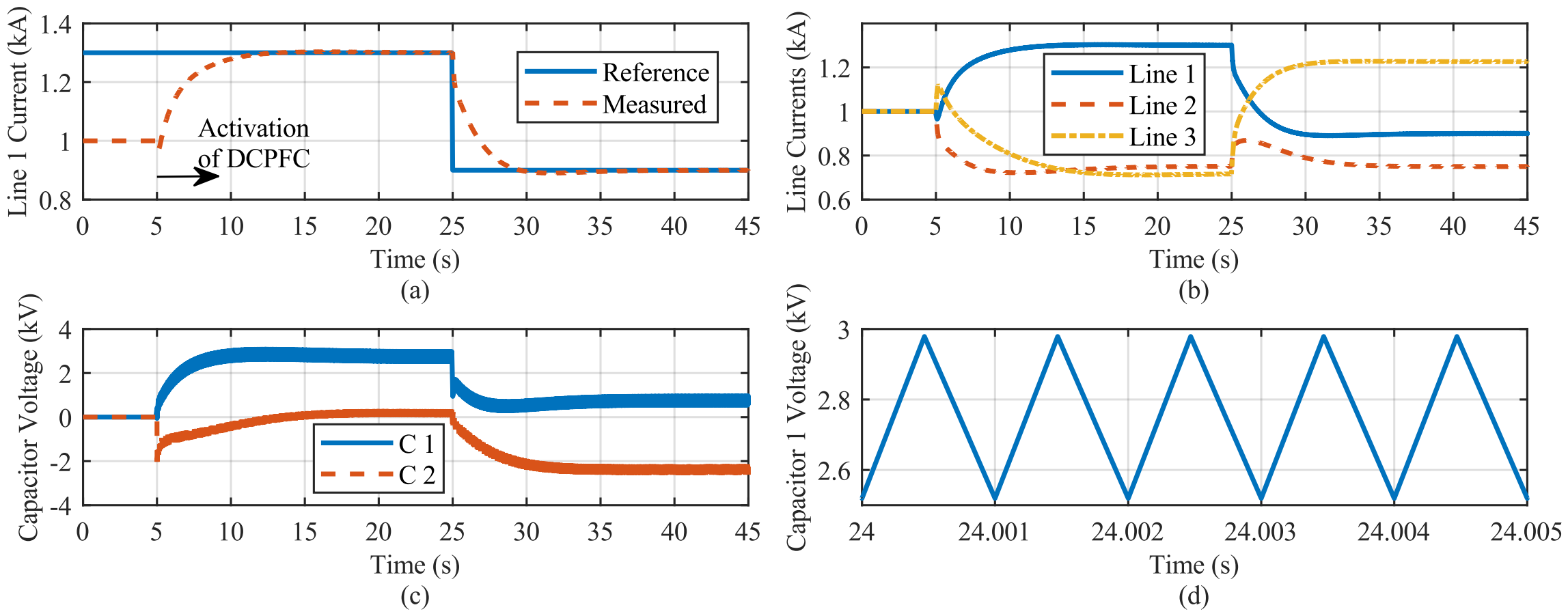

***Figure 8 – PFC operation: (a) Line 1 current and its set-point, (b) All three line currents, (c) Capacitor voltages, (d) Capacitor voltage ripple.***

Subsequently, losses associated with PFC operation are calculated based on the above-mentioned rated operating condition. The overall loss includes the switching and conduction losses of the LCS, plus only the conduction loss of the MS as they remain closed during PFC mode. The IGBT and diode models used are 4.5 kV/2 kA ABB 5SNA2000K450300 and 4.5 kV/3.23 kA Dynex DSF21545SV, respectively. The LCS is realized by 6 IGBTs and 12 diodes, whereas the MS is realized using 400 IGBTs and 800 diodes. Table 1 presents the loss breakdown and the junction temperature rise of IGBT and diode based on the system and datasheet parameters. In summary, the results show that the total loss of the three-port IM-PFCCB in the 320 kV/900 MVA system is 4697 kW, corresponding to 0.0052 p.u. and is negligible. The LCS IGBT has the largest junction temperature rise of 73.01 $^{o}$C. This is acceptable assuming an ambient temperature of 25 $^{o}$C.

***Table 1 – Component loss and temperature rise estimation***

| Component | Conduction Loss (kW) | Switching Loss (kW) | Total Loss (kW) | Component Junction Temperature Rise ($^{o}$C) |
|---|---|---|---|---|
| **LCS IGBT** | 0.925 | 13.677 | 262.836 | 73.01 |
| **LCS Diode** | 0.688 | 0.264 | 34.272 | 9.041 |
| **MS IGBT** | 2.5 | 0 | 2000 | 12.5 |
| **MS Diode** | 1.5 | 0 | 2400 | 14.25 |

# 6 Discussions

## 6.1 Topology Flexibility

As a result of the interline topology and simplified PFC circuitry, the IM-PFCCB is not only structurally scalable and modular, but also flexible enough to accommodate different applications. The next two subsections discuss possible topological modifications of the IM-PFCCB for different performance objectives.

### 6.1.1 Component/Cost Reduction

The topology shown in Fig. 2 provides maximum redundancy as each line (except Line 1 and Line $n$) can be protected by either of its two adjacent MOVs. However, it results in a relatively large number (i.e., $n-1$) and high cost of energy absorption modules. This number and cost can be reduced by only using one MOV to cover two lines it is connected to, at the expense of reduced redundancy (i.e., if this MOV degrades, then both lines are left unprotected), as illustrated by Fig. 9(a). Therefore, the minimum possible number of energy absorption modules required by IM-PFCCB is $\lceil n/2 \rceil$, where $\lceil\ \rceil$ is the ceiling operator in case $n$ is odd, to ensure that all $n$ lines are covered by an MOV.

Alternatively, it is possible to identify the more critical lines and place two MOVs adjacent to only these lines to provide redundancy. This makes the number of possible energy absorption modules in IM-PFCCB to vary between $\lceil n/2 \rceil$ and $n-1$, which depicts the customizability of IM-PFCCB and places it in between the state-of-the-art multiport DCCBs (one energy absorption module) and the conventional two-port DCCBs ($n$ energy absorption modules).

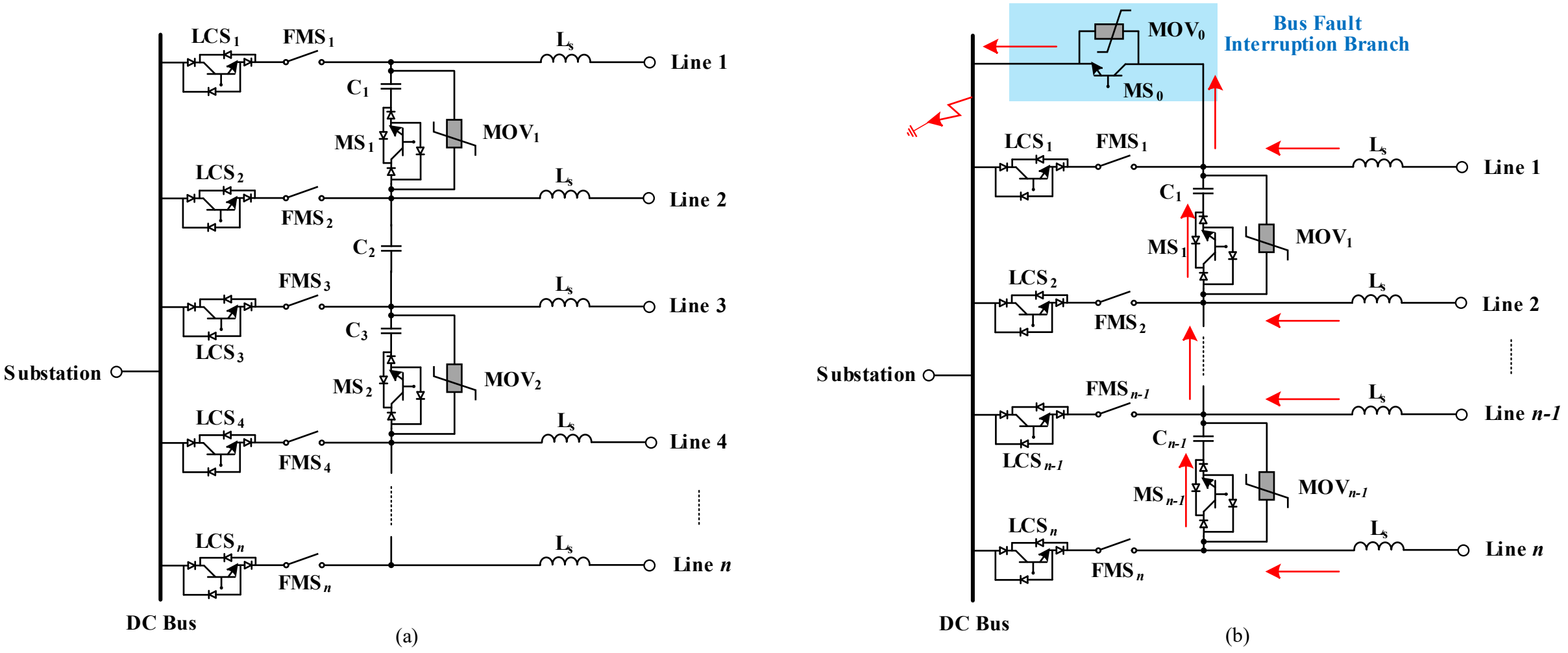

***Figure 9 – Potential topological modifications for (a) Reduced number of energy absorption modules, (b) Bus fault interruption capability (with fault current path indicated by red arrows).***

### 6.1.2 Bus Fault Interruption

The topology shown in Fig. 2 does not have inherent bus fault interruption capability. Instead, during a DC bus fault, the converter station can be isolated from the fault via (i) its AC-side breaker, (ii) a two-port DCCB installed between the converter and the bus, or (iii) a fault-blocking converter, in line with the conventional practice. Furthermore, the IM-PFCCBs installed on the remote end of Lines 1 to $n$ can be used to selectively interrupt their current flows to the local faulty bus. These options alleviate the need for the local IM-PFCCB to interrupt bus faults.

On the other hand, a possible modification of the IM-PFCCB that is able to break bus faults, at the expense of an additional energy absorption module, is shown in Fig. 9(b) [4]. In this scenario, once a bus fault is detected, all LCS and FMS are ordered to open while $MS_0$ to $MS_{n-1}$ will close. This routes the fault current into $MS_0$ path. Subsequently, $MS_0$ is turned off when FMS gains sufficient voltage withstand capability, and the fault is extinguished by $MOV_0$.

## 6.2 Application to Different HVDC Configurations

The IM-PFCCB can be deployed to different high-voltage DC (HVDC) grid configurations (e.g., monopole and bipole). As this device is installed in series between the converter and the lines, only one IM-PFCCB is required at each HVDC pole to enable power flow control and fault interruption for all lines connected to that pole.

## 6.3 Practical Design Considerations

The main practical design challenge is the trade-off among cost, complexity, and reliability. As indicated in Section 6.1, increasing the number of MOV units provides redundancy and expands the range of fault scenarios that the IM-PFCCB can interrupt, including multiple line faults and bus fault. However, this comes at the expense of higher cost and greater design complexity. Additionally, as mentioned in Section 3.2, the proposed PFC is designed for line currents with the same direction (either into or out of the substation). This limitation is the trade-off for a simple and scalable multiport implementation, since PFC topologies that can control mixed line current directions are typically more complex and less amenable to straightforward multiport extension. Another implementation challenge is the need for proper voltage and current sharing mechanisms among the series- and parallel-connected semiconductor devices to ensure reliable operation and balanced device stress.

## 6.4 Comparison with the State-of-the-Art Topologies

Table 2 provides a comparison of the proposed IM-PFCCB with the existing integrated multiport DCCB and DCPFC devices reported in the technical literature. The number of elements is expressed as a function of the number of lines connected to the converter substation (denoted as $n$). It can be observed that the main advantage of IM-PFCCB is that it provides (i) simultaneous tripping of multiple lines, (ii) backup tripping (e.g., redundancy) in case of MOV degradation and/or IGBT turn-off failure, and (iii) simple PFC control, without involving a significant component count increase compared to other integrated topologies. For example, while [4] also has simultaneous tripping capability and redundancy, its DCPFC comprises of multiple modular multilevel converter arms based on full-bridge submodules, which complicates power flow control. On the other hand, the DCPFC in [1] and [6] are simple to control, but their DCCB lacks simultaneous tripping and backup tripping capabilities. Therefore, the proposed device offers a practical alternative to the existing designs.

***Table 2 – Comparison of different integrated multiport DCCB and DCPFC devices***

| **Ref.** | **Number of Elements** | | | | **PFC Control Complexity** | **Simultaneous Tripping Capability** | **Redundancy** |
|---|---|---|---|---|---|---|---|
| | **LCS** | **FMS** | **Energy Absorption Module** | **Capacitor** | | | |
| **[1]** | $n$ | $n$ | 1 | $n-1$ | Simple | No | No |
| **[4]** | $n$ | $n$ | Variable: $\lceil n/2 \rceil$ to $n-1$ | Variable, but $\gg n$ | Complex | Yes | Yes |
| **[5]** | $4n$ | $n$ | 1 | 1 | Complex | No | No |
| **[6]** | $2n$ | $2n$ | 1 | 1 | Simple | No | No |
| **This work** | $n$ | $n$ | Variable: $\lceil n/2 \rceil$ to $n-1$ | $n-1$ | Simple | Yes | Yes |

# 7 Conclusions

This article introduces a multiport DC circuit breaker integrated with DC power flow controller, named IM-PFCCB. A key feature of the proposed topology is that the MOVs (for fault energy absorption) and capacitors (for line flow adjustment) are placed in an interline configuration. This enables, compared to the existing technologies, (i) simultaneous multiple line fault interruption, (ii) component redundancy in case of MOV degradation and/or switch failure, and (iii) simple power flow control where line currents are directly proportional to their corresponding switches' duty ratios. In addition, the topology is configurable with respect to different performance objectives, allowing bus fault interruption and cost reduction. Detailed component selection procedures are provided to facilitate the adoption of IM-PFCCB. The device's performance and efficacy are verified via EMT simulation results. Compared to the state-of-the-art topologies, IM-PFCCB achieves a well-balanced trade-off between functionality, control simplicity, and component count, making it a viable technology for MTDC grid protection and power flow control.